\documentclass{article}%
\usepackage{amsmath}
\usepackage{amsfonts}
\usepackage{amssymb}
\usepackage{graphicx}%
\begin{document}

\title{The Cosmological constant and the Wheeler-DeWitt Equation}
\author{Remo Garattini\\Universit\`{a} degli Studi di Bergamo, Facolt\`{a} di Ingegneria,\\Viale Marconi 5, 24044 Dalmine (Bergamo) Italy and\\I.N.F.N. - sezione di Milano, Milan, Italy.\\E-mail: remo.garattini@unibg.it}
\date{}
\maketitle

\begin{abstract}
We discuss how to extract information about the cosmological constant from the
Wheeler-DeWitt equation, considered as an eigenvalue of a Sturm-Liouville
problem. The equation is approximated to one loop with the help of a
variational approach with Gaussian trial wave functionals. A canonical
decomposition of modes is used to separate transverse-traceless tensors
(graviton) from ghosts and scalar. We show that no ghosts appear in the final
evaluation of the cosmological constant. A zeta function regularization is
used to handle with divergences. A renormalization procedure is introduced to
remove the infinities together with a renormalization group equation. A brief
discussion on the extension to a $f(R)$ theory is considered.

\end{abstract}

\section{Introduction}

The Friedmann-Robertson-Walker model of the universe, based on the Einstein's
field equations gives an explanation of why the Universe is in an acceleration
phase. This is supported by data observations on type I supernovae\cite{Obser}%
. Nevertheless, to obtain such an expansion we need almost 76\% of what is
known as \textit{Dark Energy}. Dark Energy is based on the following equation
of state $P=\omega\rho$ (where $P$ and $\rho$ are the pressure of the fluid
and the energy density, respectively). When $\omega<-1/3$, we are in the Dark
energy regime, while we have a transition to \textit{Phantom Energy} when
$\omega<-1$. The particular case of $\omega=-1$ corresponds to a cosmological
constant. Nevertheless, neither Dark Energy nor Phantom Energy models appear
to be satisfactory to explain the acceleration. A proposal to avoid Dark and
Phantom energy comes form the so-called modified gravity theories. In
particular, one could consider the following replacement in the
Einstein-Hilbert action\cite{f(R)} $\left(  \kappa=8\pi G\right)  $%
\begin{equation}
S=\frac{1}{2\kappa}\int d^{4}x\sqrt{-g}\!R+S^{matter}\quad\rightarrow\quad
S=\frac{1}{2\kappa}\int d^{4}x\sqrt{-g}f\left(  R\right)  +S^{matter}.
\end{equation}
It is clear that other more complicated choices could be done in place of
$f\left(  R\right)  $\cite{CF}. In particular, one could consider $f\left(
R,R_{\mu\nu}R^{\mu\nu},R_{\alpha\beta\gamma\delta}R^{\alpha\beta\gamma\delta
},\ldots\right)  $ or $f\left(  R,G\right)  $ where $G$ is the Gauss-Bonnet
invariant or any combination of these quantities\footnote{For a recent riview
on $f\left(  R\right)  $, see Refs.\cite{Faraoni,CF}, while a recent review on
the problem of $f\left(  G\right)  $ and $f\left(  R,G\right)  $ can be found
in Ref.\cite{NO,ModCosmo}.}. One of the prerogatives of a $f\left(  R\right)
$ theory is the explanation of the cosmological constant. Nevertheless,
nothing forbids to consider a more general situation where a $f\left(
R\right)  $ is combined with a cosmological constant $\Lambda_{c}$, especially
in the context of the Wheeler-DeWitt equation (WDW)\cite{De Witt}. For a
$f\left(  R\right)  =R$, one gets%
\begin{equation}
\mathcal{H=}\left(  2\kappa\right)  G_{ijkl}\pi^{ij}\pi^{kl}-\frac{\sqrt{g}%
}{2\kappa}\!{}\!\left(  \,\!^{3}R-2\Lambda_{c}\right)  =0, \label{WDW1}%
\end{equation}
where $^{3}R$ is the scalar curvature in three dimensions. The main reason to
work with a WDW equation becomes more transparent if we formally re-write the
WDW equation as\cite{Remo}%
\begin{equation}
\frac{1}{V}\frac{\int\mathcal{D}\left[  g_{ij}\right]  \Psi^{\ast}\left[
g_{ij}\right]  \int_{\Sigma}d^{3}x\hat{\Lambda}_{\Sigma}\Psi\left[
g_{ij}\right]  }{\int\mathcal{D}\left[  g_{ij}\right]  \Psi^{\ast}\left[
g_{ij}\right]  \Psi\left[  g_{ij}\right]  }=\frac{1}{V}\frac{\left\langle
\Psi\left\vert \int_{\Sigma}d^{3}x\hat{\Lambda}_{\Sigma}\right\vert
\Psi\right\rangle }{\left\langle \Psi|\Psi\right\rangle }=-\frac{\Lambda_{c}%
}{\kappa}, \label{WDW2}%
\end{equation}
where
\begin{equation}
V=\int_{\Sigma}d^{3}x\sqrt{g}%
\end{equation}
is the volume of the hypersurface $\Sigma$ and
\begin{equation}
\hat{\Lambda}_{\Sigma}=\left(  2\kappa\right)  G_{ijkl}\pi^{ij}\pi^{kl}%
-\sqrt{g}^{3}R/\left(  2\kappa\right)  .
\end{equation}
Eq.$\left(  \ref{WDW2}\right)  $ represents the Sturm-Liouville problem
associated with the cosmological constant. The related boundary conditions are
dictated by the choice of the trial wavefunctionals which, in our case are of
the Gaussian type. Different types of wavefunctionals correspond to different
boundary conditions. We can gain more information if we consider $g_{ij}%
=\bar{g}_{ij}+h_{ij},$where $\bar{g}_{ij}$ is the background metric and
$h_{ij}$ is a quantum fluctuation around the background. Thus Eq.$\left(
\ref{WDW2}\right)  $ can be expanded in terms of $h_{ij}$. Since the kinetic
part of $\hat{\Lambda}_{\Sigma}$ is quadratic in the momenta, we only need to
expand the three-scalar curvature $\int d^{3}x\sqrt{g}{}^{3}R$ up to the
quadratic order. However, to proceed with the computation, we also need an
orthogonal decomposition on the tangent space of 3-metric
deformations\cite{Vassilevich,Quad}:%

\begin{equation}
h_{ij}=\frac{1}{3}\left(  \sigma+2\nabla\cdot\xi\right)  g_{ij}+\left(
L\xi\right)  _{ij}+h_{ij}^{\bot}. \label{p21a}%
\end{equation}
The operator $L$ maps $\xi_{i}$ into symmetric tracefree tensors
\begin{equation}
\left(  L\xi\right)  _{ij}=\nabla_{i}\xi_{j}+\nabla_{j}\xi_{i}-\frac{2}%
{3}g_{ij}\left(  \nabla\cdot\xi\right)  ,
\end{equation}
$h_{ij}^{\bot}$ is the traceless-transverse component of the perturbation
(TT), namely
\begin{equation}
g^{ij}h_{ij}^{\bot}=0,\qquad\nabla^{i}h_{ij}^{\bot}=0
\end{equation}
and $h$ is the trace of $h_{ij}$. It is immediate to recognize that the trace
element $\sigma=h-2\left(  \nabla\cdot\xi\right)  $ is gauge invariant. If we
perform the same decomposition also on the momentum $\pi^{ij}$, up to second
order Eq.$\left(  \ref{WDW2}\right)  $ becomes%
\begin{equation}
\frac{1}{V}\frac{\left\langle \Psi\left\vert \int_{\Sigma}d^{3}x\left[
\hat{\Lambda}_{\Sigma}^{\bot}+\hat{\Lambda}_{\Sigma}^{\xi}+\hat{\Lambda
}_{\Sigma}^{\sigma}\right]  ^{\left(  2\right)  }\right\vert \Psi\right\rangle
}{\left\langle \Psi|\Psi\right\rangle }=-\frac{\Lambda_{c}}{\kappa}\Psi\left[
g_{ij}\right]  . \label{lambda0_2}%
\end{equation}
Concerning the measure appearing in Eq.$\left(  \ref{WDW2}\right)  $, we have
to note that the decomposition $\left(  \ref{p21a}\right)  $ induces the
following transformation on the functional measure $\mathcal{D}h_{ij}%
\rightarrow\mathcal{D}h_{ij}^{\bot}\mathcal{D}\xi_{i}\mathcal{D}\sigma J_{1}$,
where the Jacobian related to the gauge vector variable $\xi_{i}$ is%
\begin{equation}
J_{1}=\left[  \det\left(  \bigtriangleup g^{ij}+\frac{1}{3}\nabla^{i}%
\nabla^{j}-R^{ij}\right)  \right]  ^{\frac{1}{2}}.
\end{equation}
This is nothing but the famous Faddev-Popov determinant. It becomes more
transparent if $\xi_{a}$ is further decomposed into a transverse part $\xi
_{a}^{T}$ with $\nabla^{a}\xi_{a}^{T}=0$ and a longitudinal part $\xi
_{a}^{\parallel}$ with $\xi_{a}^{\parallel}=$ $\nabla_{a}\psi$, then $J_{1}$
can be expressed by an upper triangular matrix for certain backgrounds (e.g.
Schwarzschild in three dimensions). It is immediate to recognize that for an
Einstein space in any dimension, cross terms vanish and $J_{1}$ can be
expressed by a block diagonal matrix. Since $\det AB=\det A\det B$, the
functional measure $\mathcal{D}h_{ij}$ factorizes into%
\begin{equation}
\mathcal{D}h_{ij}=\left(  \det\bigtriangleup_{V}^{T}\right)  ^{\frac{1}{2}%
}\left(  \det\left[  \frac{2}{3}\bigtriangleup^{2}+\nabla_{i}R^{ij}\nabla
_{j}\right]  \right)  ^{\frac{1}{2}}\mathcal{D}h_{ij}^{\bot}\mathcal{D}\xi
^{T}\mathcal{D}\psi
\end{equation}
with $\left(  \bigtriangleup_{V}^{ij}\right)  ^{T}=\bigtriangleup
g^{ij}-R^{ij}$ acting on transverse vectors, which is the Faddeev-Popov
determinant. In writing the functional measure $\mathcal{D}h_{ij}$, we have
here ignored the appearance of a multiplicative anomaly\cite{EVZ}. Thus the
inner product can be written as%
\begin{equation}
\int\mathcal{D}h_{ij}^{\bot}\mathcal{D}\xi^{T}\mathcal{D}\sigma\Psi^{\ast
}\left[  h_{ij}^{\bot}\right]  \Psi^{\ast}\left[  \xi^{T}\right]  \Psi^{\ast
}\left[  \sigma\right]  \Psi\left[  h_{ij}^{\bot}\right]  \Psi\left[  \xi
^{T}\right]  \Psi\left[  \sigma\right]  \left(  \det\bigtriangleup_{V}%
^{T}\right)  ^{\frac{1}{2}}\left(  \det\left[  \frac{2}{3}\bigtriangleup
^{2}+\nabla_{i}R^{ij}\nabla_{j}\right]  \right)  ^{\frac{1}{2}}.
\end{equation}
Nevertheless, since there is no interaction between ghost fields and the other
components of the perturbation at this level of approximation, the Jacobian
appearing in the numerator and in the denominator simplify. The reason can be
found in terms of connected and disconnected terms. The disconnected terms
appear in the Faddeev-Popov determinant and these ones are not linked by the
Gaussian integration. This means that disconnected terms in the numerator and
the same ones appearing in the denominator cancel out. Therefore, Eq.$\left(
\ref{lambda0_2}\right)  $ factorizes into three pieces. The piece containing
$\hat{\Lambda}_{\Sigma}^{\bot}$ is the contribution of the
transverse-traceless tensors (TT): essentially is the graviton contribution
representing true physical degrees of freedom. Regarding the vector term
$\hat{\Lambda}_{\Sigma}^{T}$, we observe that under the action of
infinitesimal diffeomorphism generated by a vector field $\epsilon_{i}$, the
components of $\left(  \ref{p21a}\right)  $ transform as
follows\cite{Vassilevich}%
\begin{equation}
\xi_{j}\longrightarrow\xi_{j}+\epsilon_{j},\qquad h\longrightarrow
h+2\nabla\cdot\xi,\qquad h_{ij}^{\bot}\longrightarrow h_{ij}^{\bot}.
\end{equation}
The Killing vectors satisfying the condition $\nabla_{i}\xi_{j}+\nabla_{j}%
\xi_{i}=0,$ do not change $h_{ij}$, and thus should be excluded from the gauge
group. All other diffeomorphisms act on $h_{ij}$ nontrivially. We need to fix
the residual gauge freedom on the vector $\xi_{i}$. The simplest choice is
$\xi_{i}=0.$This new gauge fixing produces the same Faddeev-Popov determinant
connected to the Jacobian $J_{1}$ and therefore will not contribute to the
final value. We are left with%
\begin{equation}
\frac{1}{V}\frac{\left\langle \Psi^{\bot}\left\vert \int_{\Sigma}d^{3}x\left[
\hat{\Lambda}_{\Sigma}^{\bot}\right]  ^{\left(  2\right)  }\right\vert
\Psi^{\bot}\right\rangle }{\left\langle \Psi^{\bot}|\Psi^{\bot}\right\rangle
}+\frac{1}{V}\frac{\left\langle \Psi^{\sigma}\left\vert \int_{\Sigma}%
d^{3}x\left[  \hat{\Lambda}_{\Sigma}^{\sigma}\right]  ^{\left(  2\right)
}\right\vert \Psi^{\sigma}\right\rangle }{\left\langle \Psi^{\sigma}%
|\Psi^{\sigma}\right\rangle }=-\frac{\Lambda_{c}}{\kappa}\Psi\left[
g_{ij}\right]  . \label{lambda0_2a}%
\end{equation}
Note that in the expansion of $\int_{\Sigma}d^{3}x\sqrt{g}{}R$ to second
order, a coupling term between the TT component and scalar one remains.
However, the Gaussian integration does not allow such a mixing which has to be
introduced with an appropriate wave functional. Extracting the TT tensor
contribution from Eq.$\left(  \ref{WDW2}\right)  $ approximated to second
order in perturbation of the spatial part of the metric into a background term
$\bar{g}_{ij}$, and a perturbation $h_{ij}$, we get%
\begin{equation}
\hat{\Lambda}_{\Sigma}^{\bot}=\frac{1}{4V}\int_{\Sigma}d^{3}x\sqrt{\bar{g}%
}G^{ijkl}\left[  \left(  2\kappa\right)  K^{-1\bot}\left(  x,x\right)
_{ijkl}+\frac{1}{\left(  2\kappa\right)  }\!{}\left(  \tilde{\bigtriangleup
}_{L\!}\right)  _{j}^{a}K^{\bot}\left(  x,x\right)  _{iakl}\right]  ,
\label{p22}%
\end{equation}
where%
\begin{equation}
\left(  \tilde{\bigtriangleup}_{L\!}\!{}h^{\bot}\right)  _{ij}=\left(
\bigtriangleup_{L\!}\!{}h^{\bot}\right)  _{ij}-4R{}_{i}^{k}\!{}h_{kj}^{\bot
}+\text{ }^{3}R{}\!{}h_{ij}^{\bot} \label{M Lichn}%
\end{equation}
is the modified Lichnerowicz operator and $\bigtriangleup_{L}$is the
Lichnerowicz operator defined by%
\begin{equation}
\left(  \bigtriangleup_{L}h\right)  _{ij}=\bigtriangleup h_{ij}-2R_{ikjl}%
h^{kl}+R_{ik}h_{j}^{k}+R_{jk}h_{i}^{k}\qquad\bigtriangleup=-\nabla^{a}%
\nabla_{a}. \label{DeltaL}%
\end{equation}
$G^{ijkl}$ represents the inverse DeWitt metric and all indices run from one
to three. Note that the term%
\begin{equation}
-4R{}_{i}^{k}\!{}h_{kj}^{\bot}+^{3}R{}\!{}h_{ij}^{\bot}%
\end{equation}
disappears in four dimensions. The propagator $K^{\bot}\left(  x,x\right)
_{iakl}$ can be represented as
\begin{equation}
K^{\bot}\left(  \overrightarrow{x},\overrightarrow{y}\right)  _{iakl}%
=\sum_{\tau}\frac{h_{ia}^{\left(  \tau\right)  \bot}\left(  \overrightarrow
{x}\right)  h_{kl}^{\left(  \tau\right)  \bot}\left(  \overrightarrow
{y}\right)  }{2\lambda\left(  \tau\right)  }, \label{proptt}%
\end{equation}
where $h_{ia}^{\left(  \tau\right)  \bot}\left(  \overrightarrow{x}\right)  $
are the eigenfunctions of $\tilde{\bigtriangleup}_{L\!}$. $\tau$ denotes a
complete set of indices and $\lambda\left(  \tau\right)  $ are a set of
variational parameters to be determined by the minimization of Eq.$\left(
\ref{p22}\right)  $. The expectation value of $\hat{\Lambda}_{\Sigma}^{\bot} $
is easily obtained by inserting the form of the propagator into Eq.$\left(
\ref{p22}\right)  $ and minimizing with respect to the variational function
$\lambda\left(  \tau\right)  $. Thus the total one loop energy density for TT
tensors becomes%
\begin{equation}
\frac{\Lambda}{8\pi G}=-\frac{1}{2}\sum_{\tau}\left[  \sqrt{\omega_{1}%
^{2}\left(  \tau\right)  }+\sqrt{\omega_{2}^{2}\left(  \tau\right)  }\right]
. \label{1loop}%
\end{equation}
The above expression makes sense only for $\omega_{i}^{2}\left(  \tau\right)
>0$, where $\omega_{i}$ are the eigenvalues of $\tilde{\bigtriangleup}_{L\!}$.
Concerning the scalar contribution of Eq.$\left(  \ref{lambda0_2a}\right)  $,
in Ref.\cite{Remo1} has been proved that the cosmological constant
contribution is%
\begin{equation}
\frac{\Lambda^{\sigma}}{8\pi G}=\frac{1}{4}\sqrt{\frac{2}{3}}\sum_{\tau
}\left[  \sqrt{\omega^{2}\left(  \tau\right)  }\right]  , \label{1loops}%
\end{equation}
where $\omega\left(  \tau\right)  $ is the eigenvalue of the scalar part of
the perturbation. In the next section, we will explictly evaluate Eqs.$\left(
\ref{1loop},\ref{1loops}\right)  $ for a specific background.

\section{One loop energy Regularization and Renormalization for a $f\left(
R\right)  =R$ theory}

If we consider a background of the form%
\begin{equation}
ds^{2}=-N^{2}\left(  r\right)  dt^{2}+\frac{dr^{2}}{1-\frac{b\left(  r\right)
}{r}}+r^{2}\left(  d\theta^{2}+\sin^{2}\theta d\phi^{2}\right)  ,
\label{metric}%
\end{equation}
then, with the help of Regge and Wheeler representation\cite{Regge Wheeler},
$\left(  \tilde{\bigtriangleup}_{L\!}\!{}h^{\bot}\right)  _{ij}$ can be
reduced to%
\begin{equation}
\left[  -\frac{d^{2}}{dx^{2}}+\frac{l\left(  l+1\right)  }{r^{2}}+m_{i}%
^{2}\left(  r\right)  \right]  f_{i}\left(  x\right)  =\omega_{i,l}^{2}%
f_{i}\left(  x\right)  \quad i=1,2\quad, \label{p34}%
\end{equation}
where we have used reduced fields of the form $f_{i}\left(  x\right)
=F_{i}\left(  x\right)  /r$ and where we have defined two r-dependent
effective masses $m_{1}^{2}\left(  r\right)  $ and $m_{2}^{2}\left(  r\right)
$%
\begin{equation}
\left\{
\begin{array}
[c]{c}%
m_{1}^{2}\left(  r\right)  =\frac{6}{r^{2}}\left(  1-\frac{b\left(  r\right)
}{r}\right)  +\frac{3}{2r^{2}}b^{\prime}\left(  r\right)  -\frac{3}{2r^{3}%
}b\left(  r\right) \\
\\
m_{2}^{2}\left(  r\right)  =\frac{6}{r^{2}}\left(  1-\frac{b\left(  r\right)
}{r}\right)  +\frac{1}{2r^{2}}b^{\prime}\left(  r\right)  +\frac{3}{2r^{3}%
}b\left(  r\right)
\end{array}
\right.  \quad\left(  r\equiv r\left(  x\right)  \right)  . \label{masses}%
\end{equation}
In order to use the WKB approximation, from Eq.$\left(  \ref{p34}\right)  $ we
can extract two r-dependent radial wave numbers%
\begin{equation}
k_{i}^{2}\left(  r,l,\omega_{i,nl}\right)  =\omega_{i,nl}^{2}-\frac{l\left(
l+1\right)  }{r^{2}}-m_{i}^{2}\left(  r\right)  \quad i=1,2\quad. \label{kTT}%
\end{equation}
When $b\left(  r\right)  =r_{t}=2MG$, the effective masses can be approximated
in the range where $r\in\left[  r_{t},5r_{t}/2\right]  $ with $m_{1}%
^{2}\left(  r\right)  =-m_{2}^{2}\left(  r\right)  =m_{0}^{2}\left(  r\right)
$. Such a restriction comes from the fact that the effective masses, in this
range, represent short distance contribution. Indeed, we expect to receive
large contribution from quantum fluctuations at short distances. It is now
possible to explicitly evaluate Eq.$\left(  \ref{1loop}\right)  $ in terms of
the effective mass. To further proceed we use the W.K.B. method used by `t
Hooft in the brick wall problem\cite{tHooft} and we count the number of modes
with frequency less than $\omega_{i}$, $i=1,2$. This is given approximately by%
\begin{equation}
\tilde{g}\left(  \omega_{i}\right)  =\int_{0}^{l_{\max}}\nu_{i}\left(
l,\omega_{i}\right)  \left(  2l+1\right)  dl, \label{p41}%
\end{equation}
where $\nu_{i}\left(  l,\omega_{i}\right)  $, $i=1,2$ is the number of nodes
in the mode with $\left(  l,\omega_{i}\right)  $, such that $\left(  r\equiv
r\left(  x\right)  \right)  $
\begin{equation}
\nu_{i}\left(  l,\omega_{i}\right)  =\frac{1}{\pi}\int_{-\infty}^{+\infty
}dx\sqrt{k_{i}^{2}\left(  r,l,\omega_{i}\right)  }. \label{p42}%
\end{equation}
Here it is understood that the integration with respect to $x$ and $l_{\max} $
is taken over those values which satisfy $k_{i}^{2}\left(  r,l,\omega
_{i}\right)  \geq0,$ $i=1,2$. With the help of Eqs.$\left(  \ref{p41}%
,\ref{p42}\right)  $, Eq.$\left(  \ref{1loop}\right)  $ becomes%
\begin{equation}
\frac{\Lambda}{8\pi G}=-\frac{1}{\pi}\sum_{i=1}^{2}\int_{0}^{+\infty}%
\omega_{i}\frac{d\tilde{g}\left(  \omega_{i}\right)  }{d\omega_{i}}d\omega
_{i}. \label{tot1loop}%
\end{equation}
This is the graviton contribution to the induced cosmological constant to one
loop. The explicit evaluation of Eq.$\left(  \ref{tot1loop}\right)  $ gives%
\begin{equation}
\frac{\Lambda}{8\pi G}=\rho_{1}+\rho_{2}=-\frac{1}{4\pi^{2}}\sum_{i=1}^{2}%
\int_{\sqrt{m_{i}^{2}\left(  r\right)  }}^{+\infty}\omega_{i}^{2}\sqrt
{\omega_{i}^{2}-m_{i}^{2}\left(  r\right)  }d\omega_{i}, \label{tote1loop}%
\end{equation}
where we have included an additional $4\pi$ coming from the angular
integration. The use of the zeta function regularization method to compute the
energy densities $\rho_{1}$ and $\rho_{2}$ leads to%
\begin{equation}
\rho_{i}\left(  \varepsilon\right)  =\frac{m_{i}^{4}\left(  r\right)  }%
{64\pi^{2}}\left[  \frac{1}{\varepsilon}+\ln\left(  \frac{4\mu^{2}}{m_{i}%
^{2}\left(  r\right)  \sqrt{e}}\right)  \right]  \quad i=1,2\quad,
\label{zeta1}%
\end{equation}
where we have introduced the additional mass parameter $\mu$ in order to
restore the correct dimension for the regularized quantities. Such an
arbitrary mass scale emerges unavoidably in any regularization scheme. The
renormalization is performed via the absorption of the divergent part into the
re-definition of the bare classical constant $\Lambda$, namely $\Lambda
\rightarrow\Lambda_{0}+\Lambda^{div}$. The remaining finite value for the
cosmological constant reads%
\begin{equation}
\frac{\Lambda_{0}}{8\pi G}=\left(  \rho_{1}\left(  \mu\right)  +\rho
_{2}\left(  \mu\right)  \right)  =\rho_{eff}^{TT}\left(  \mu,r\right)  ,
\label{lambda0}%
\end{equation}
where $\rho_{i}\left(  \mu\right)  $ has the same form of $\rho_{1}\left(
\varepsilon\right)  $ but without the divergence. The quantity in Eq.$\left(
\ref{lambda0}\right)  $ depends on the arbitrary mass scale $\mu.$ It is
appropriate to use the renormalization group equation to eliminate such a
dependence. To this aim, we impose that\cite{RGeq}%

\begin{equation}
\frac{1}{8\pi G}\mu\frac{\partial\Lambda_{0}\left(  \mu\right)  }{\partial\mu
}=\mu\frac{d}{d\mu}\rho_{eff}^{TT}\left(  \mu,r\right)  . \label{rg}%
\end{equation}
Solving it we find that the renormalized constant $\Lambda_{0}$ should be
treated as a running one in the sense that it varies provided that the scale
$\mu$ is changing%

\begin{equation}
\frac{\Lambda_{0}\left(  \mu,r\right)  }{8\pi G}=\frac{\Lambda_{0}\left(
\mu_{0},r\right)  }{8\pi G}+\frac{m_{0}^{4}\left(  r\right)  }{16\pi^{2}}%
\ln\frac{\mu}{\mu_{0}}.\label{lambdamu}%
\end{equation}
Substituting Eq.$\left(  \ref{lambdamu}\right)  $ into Eq.$\left(
\ref{lambda0}\right)  $ we find%
\begin{equation}
\frac{\Lambda_{0}\left(  \mu_{0},r\right)  }{8\pi G}=-\frac{1}{32\pi^{2}%
}\left\{  m_{0}^{4}\left(  r\right)  \left[  \ln\left(  \frac{m_{0}^{2}\left(
r\right)  \sqrt{e}}{4\mu_{0}^{2}}\right)  \right]  \right\}
.\label{lambdamu0}%
\end{equation}
If we go back and look at Eq.$\left(  \ref{WDW2}\right)  $, we note that what
we have actually computed is the opposite of an effective potential (better an
effective energy). Therefore, we expect to find physically acceptable
solutions in proximity of the extrema. We find that Eq.$\left(
\ref{lambdamu0}\right)  $ has an extremum when%
\begin{equation}
\frac{1}{e}=\frac{m_{0}^{2}\left(  r\right)  }{4\mu_{0}^{2}}\qquad
\Longrightarrow\qquad\frac{\bar{\Lambda}_{0}\left(  \mu_{0},\bar{r}\right)
}{8\pi G}=\frac{m_{0}^{4}\left(  \bar{r}\right)  }{64\pi^{2}}=\frac{\mu
_{0}^{4}}{4\pi^{2}e^{2}}.\label{LambdansM}%
\end{equation}
Actually $\bar{\Lambda}_{0}\left(  \mu_{0},\bar{r}\right)  $ is a maximum,
corresponding to a minimum of the effective energy. Note also that there
exists another extremum when%
\begin{equation}
m_{0}^{4}\left(  r\right)  =0\qquad\Longrightarrow\qquad M=0.
\end{equation}
This solution corresponds to Minkowski space, producing no effect on the
vacuum. For this reason it will be discarded. On the other hand, the effect of
the gravitational fluctuations is to shift the minimum of the effective energy
away from the flat solution leading to an induced cosmological constant. If we
apply the same procedure to the scalar part of the perturbation, we find that
the only consistent solution is that $\Lambda^{\sigma}=0$. Therefore, the
whole contribution is due to the physical degrees of freedom: the
graviton\cite{GriKos}. Plugging Eq.$\left(  \ref{LambdansM}\right)  $ into
Eq.$\left(  \ref{lambdamu}\right)  $, we find%
\begin{equation}
\frac{\Lambda_{0}\left(  \mu,\bar{r}\right)  }{8\pi G}=\frac{\bar{\Lambda}%
_{0}\left(  \mu_{0},\bar{r}\right)  }{8\pi G}+\frac{m_{0}^{4}\left(  \bar
{r}\right)  }{16\pi^{2}}\ln\frac{\mu}{\mu_{0}}=\frac{m_{0}^{4}\left(  \bar
{r}\right)  }{64\pi^{2}}\left(  1+4\ln\frac{\mu}{\mu_{0}}\right)  .
\end{equation}
If we set $\mu_{0}=m_{P}$, where $m_{P}$ is the Planck mass, we can find that%
\begin{equation}
\frac{\Lambda_{0}\left(  \tilde{\mu},r\right)  }{8\pi G}=0\qquad
\text{\textrm{when\qquad}}\tilde{\mu}=\exp\left(  -\frac{1}{2}\right)  \mu
_{0}.
\end{equation}
Nevertheless, $\tilde{\mu}$ is of the order of the Planck mass again, but
unfortunately is a scale which is very far from the nowadays observations.
However, it is interesting to note that this approach can be generalized by
replacing the scalar curvature $R$ with a generic function of $R$. Although a
$f\left(  R\right)  $ theory does not need a cosmological constant, rather it
should explain it, we shall consider the following Lagrangian density
describing a generic $f(R)$ theory of gravity
\begin{equation}
\mathcal{L}=\sqrt{-g}\left(  f\left(  R\right)  -2\Lambda\right)
,\qquad{with}\;f^{\prime\prime}\neq0,\label{lag}%
\end{equation}
where $f\left(  R\right)  $ is an arbitrary smooth function of the scalar
curvature and primes denote differentiation with respect to the scalar
curvature. A cosmological term is added also in this case for the sake of
generality, because in any case, Eq.$\left(  \ref{lag}\right)  $ represents
the most general lagrangian to examine. Obviously $f^{\prime\prime}=0$
corresponds to GR.\cite{Querella}. The semi-classical procedure followed in
this work relies heavily on the formalism outlined in Refs.\cite{Remo1,CG}.
The main effect of this replacement is that at the scale $\mu_{0}$, we have a
shift of the old induced cosmological constant into%
\begin{equation}
\frac{\Lambda_{0}^{\prime}\left(  \mu_{0},r\right)  }{8\pi G}=\frac{1}%
{\sqrt{h\left(  R\right)  }}\left[  \frac{\Lambda_{0}\left(  \mu_{0},r\right)
}{8\pi G}+\frac{1}{16\pi GV}\int_{\Sigma}d^{3}x\sqrt{g}\frac{Rf^{\prime
}\left(  R\right)  -f\left(  R\right)  }{f^{\prime}\left(  R\right)  }\right]
,\label{Lambdaf(R)}%
\end{equation}
where $V$ is the volume of the system. Note that when $f\left(  R\right)  =R$,
consistently it is $h\left(  R\right)  =1$ with%
\begin{equation}
h\left(  R\right)  =1+\frac{2\left[  f^{\prime}\left(  R\right)  -1\right]
}{f^{\prime}\left(  R\right)  }\label{h(R)}%
\end{equation}
We can always choose the form of $f\left(  R\right)  $ in such a way
$\Lambda_{0}\left(  \mu_{0},r\right)  $. This implies%
\begin{equation}
\frac{\Lambda_{0}^{\prime}\left(  \mu_{0},r\right)  }{8\pi G}=\frac{1}%
{\sqrt{h\left(  R\right)  }}\frac{1}{16\pi GV}\int_{\Sigma}d^{3}x\sqrt{g}%
\frac{Rf^{\prime}\left(  R\right)  -f\left(  R\right)  }{f^{\prime}\left(
R\right)  }.
\end{equation}
A comment is in order. We have found that our calculation is in agreement with
Ref.\cite{GriKos}, where only the graviton contribution is fundamental. Note
also the absence of a Faddeev-Popov determinant. This is in agreement with
Ref.\cite{GriKos} but also with Ref.\cite{Vassilevich}, where the
Faddeev-Popov determinant appears when perturbations of the shift vectors are
considered. The second comment regards our one loop computation which is
deeply different form the one loop computation of Refs.\cite{CENOZ,MS}, where
the analysis has been done expanding directly $f\left(  R\right)  $. In our
case, the expansion involves only the three dimensional scalar curvature. Note
that with the metric $\left(  \ref{metric}\right)  $ and the effective masses
$\left(  \ref{masses}\right)  $, in principle, we can examine every
spherically symmetric metric. Note also the absence of boundary terms in the
evaluation of the induced cosmological constant.

\end{document}